\documentclass[aip,pof,graphicx]{revtex4}

\usepackage{amsmath, amssymb, epsf,rotate}
\usepackage{graphicx}
\newcommand{\e}{\varepsilon}

 \newcommand{\drp}[2]{\frac{\partial #1}{\partial #2}}

\newcommand{\be}{\begin{equation}}
\newcommand{\ee}{\end{equation}}
\newcommand{\hi}{h_{\infty}}

\begin{document}

\title{Liquid meniscus friction on a wet plate: Bubbles, lamellae and foams}

\author{Isabelle Cantat}
\email[]{isabelle.cantat@univ-rennes1.fr}
\affiliation{Institut de Physique de Rennes, UMR CNRS 6251, Universit\'e Rennes 1,  B\^atiment 11A, Campus de Beaulieu, 35042 Rennes cedex, France.}

\date{\today}

\begin{abstract}
Many microfluidics devices, coating processes or diphasic flows involve the motion of a liquid meniscus on a wet wall. This motion induces  a specific viscous  force, that exhibits a non-linear dependency in the meniscus velocity.   We propose a review of the theoretical and experimental work made on this viscous force, for simple interfacial properties. The interface is indeed assumed either perfectly compressible (mobile interface) or perfectly incompressible (rigid interface).  We show that, in the second case, the viscous force exerted by the wall on the meniscus is a combination of two power laws, scaling like $Ca^{1/3}$ and $Ca^{2/3}$, with $Ca$ the capillary number. We provide a prediction for the stress exerted on a foam  sliding on a wet solid and compare it with experimental data, for the incompressible case. 
\end{abstract}

\maketitle

Many industrial or natural situations involve the motion of  liquid menisci on a wet solid. For enhanced oil recovery \cite{hirasaki85, rossen90a, kornev99}, or soil remediation \cite{chowdiah98}, liquid foams are pushed into porous media to displace oil or pollutants;  surfactant lamellae are produced in the lung and move in the bronchial tubes \cite{grotberg01,howell00}; innovative set-ups are developed  to control  bubbles motion  in micro-channels for lab-on-a-chip applications \cite{baroud10};  pulling a solid out of a liquid bath is a common way used in industry for coating \cite{quere99}. 

\begin{figure}[h]
\centerline{\includegraphics[width=14cm]{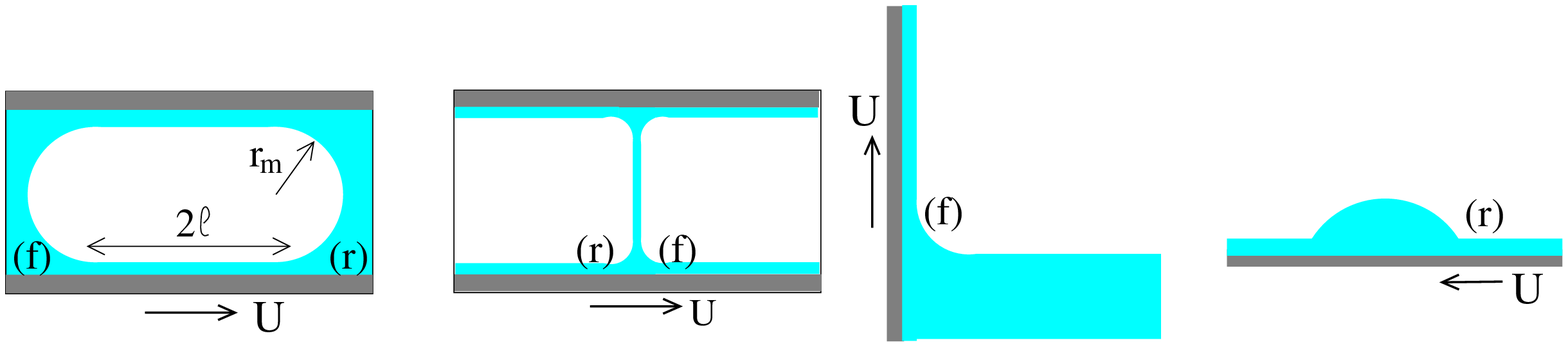}}
\caption{Different diphasic flows involving the motion of a meniscus on a wet wall.  From left to right: a single bubble in a tube (Bretherton's problem \cite{bretherton61}) or in a Hele-Shaw cell; a lamella in a tube; a plate or a fiber \cite{quere99} pulled out of a liquid bath (Landau-Levich-Derjaguin (LLD) problem \cite{landau42,derjaguin33}); a droplet spreading on a wet plate (Tanner's problem \cite{tanner79}). $U$ is the velocity of the solid in the frame of the meniscus. The front and rear menisci are denoted respectively by (f) and (r). Each meniscus is connected to a wetting film of length $\ell$.}
\label{fig_configuration}
\end{figure}

As depicted in Fig. \ref{fig_configuration}, the same local flow is observed close to the wall in all these examples.  
While the meniscus moves on the wall, it deposits a thin wetting film. Its thickness, of the order of few tens of microns, depends on the meniscus velocity through the capillary number $Ca= \eta U/\gamma$ that compares the viscous stress and the capillary pressure, through the values of  the solution viscosity $\eta$, the surface tension $\gamma$ and the meniscus velocity $U$. This property, that makes possible a finely tunable coating, has been widely investigated \cite{quere99}. 
In contrast, the force required to pull a meniscus at a given velocity on a wall is much less understood, and plays a crucial role for the motion of  confined bubbles or confined foams. This force is governed by the velocity gradients in the liquid phase. As the liquid phase is confined into the small gaps between the bubbles and the wall, these gradients, and consequently the viscous force, are strongly enhanced and depend on the dynamical shapes of the meniscus and of the wetting film.
As a direct consequence, pushing a foam in a tube requires a pressure drop that is much higher than the one needed to push the surfactant solution in the same tube at the same mean velocity \cite{hohler05,livre_mousse}.

The theoretical prediction of the shape of the deposited film has been proposed by Landau and Levich \cite{landau42} and by Derjaguin \cite{derjaguin43} (LLD) in the limit of small capillary numbers.  Bretherton \cite{bretherton61} and Park and Homsy \cite{park84} adapted  this calculation to the  case of a confined bubble and predicted the pressure drop associated with the bubble motion. Experimental results mainly report on  the film thickness deposited by long  bubbles  in cylindrical tubes \cite{fairbrother35,taylor61,bretherton61,chen86,schwartz86,aussillous00,emile12} or on solids pulled out of a bath (\cite{quere99} and references therein, \cite{chan12}), for a typical range of capillary number $[10^{-6} ; 10^{-3}]$. The viscous force acting on the meniscus, or the pressure drop, have been measured  for different systems including bubbles separated by liquid slugs \cite{cubaud04,fuerstman07} or  lamella \cite{hirasaki85,cantat04,terriac06,dollet10,bazilevsky12} moving in tubes, and foams in 2D \cite{raufaste09}  or
  3D geometries \cite{denkov05,marze08}.
Numerical simulations have extended the predictions for the meniscus induced  pressure drop  to higher capillary numbers or to more complex geometries \cite{reinelt85,reinelt87,ratulowski89,martinez90,giavedoni97,heil01,hazel02,saugey06,fujioka08}. 

Lamella and foams are stabilized by surfactants that induce a specific stress  at the interface. 
If the area $A$ of an interface element increases fast enough, the surfactant concentration decreases and the surface tension increases, thus acting against the  extension. The dilatational modulus of the interface, defined as $E=(1/A)\partial \gamma /\partial A$, is a measure of this phenomenon.  Moreover the 
intrinsic viscosity $\eta_s$ of the surfactant monolayer induces an additional interfacial stress, that also acts against the extension. In the limit of high dilatational modulus or high interfacial viscosity,  the interface behaves as an incompressible surface.

In this paper, we discuss the motion of a meniscus on a wet wall, in the ideal cases of incompressible interfaces or interfaces with uniform tangential stress. The results obtained in these limiting cases  constitute important references for models with more realistic interfacial properties. 
Section \ref{film_shape} presents classical calculations for the dynamical meniscus shape, from which the viscous forces are deduced in 
Section \ref{section_force}, as well as the validity range of the ideal models. An original derivation of the pressure drop for a single bubble is proposed in Section \ref{section_pressure} and compared with Bretherthon's calculation. Finally, adapting  Denkov's approach \cite{denkov05}, these results are used in Section \ref{filmfoam} to derive predictions for the foam/wall tangential stress as a function of the bubble size and liquid fraction, for stress-free or incompressible interfaces. For incompressible interfaces, we show that the stress is not a power law of the capillary number as usually assumed, but rather a sum of two contributions scaling as $Ca^{1/3}$ and $Ca^{2/3}$, in good agreement with experimental results. 

\section{Dynamical meniscus shape}
\label{film_shape}

\subsection{Resolution of the lubrication problem}
\label{film_prof}

\begin{figure}
\centerline{\includegraphics[width=8cm]{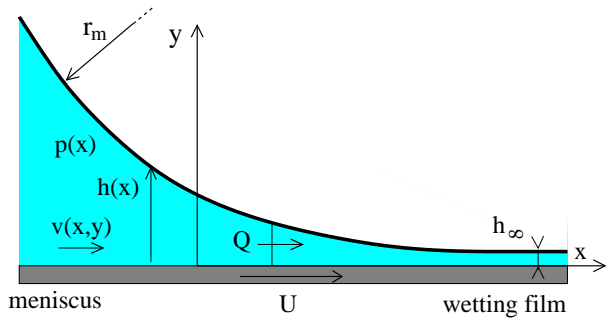} }
\caption{Schematic view of a liquid meniscus on a solid that moves in the $x$-direction at velocity $U$. The thickness profile is $h(x)$, the pressure is $p(x)$ and the x-component of the velocity in the meniscus frame is $v(x,y)$. The radius of curvature of the meniscus tends to $r_m$ for $x\to -\infty$ (meniscus side) and the film thickness tends to $\hi$  for $x\to \infty$ (wetting film side). The plate velocity $U>0$ corresponds to a front meniscus (f) and $U<0$ to a rear meniscus (r). The problem is invariant in the $z$ direction.
}
\label{fig_notation}
\end{figure}

We consider a liquid meniscus moving at constant velocity in the $x$ direction on a solid plate, in steady state. We assume an invariance in the $z$ direction  (see the convention of orientation on Fig. \ref{fig_notation}) and we neglect inertial and gravitational effects. In the meniscus frame, a positive plate velocity $U>0$ corresponds to a plate pulled out of a bath as in the Landau Levich problem, to the front end of the bubble as in the Bretherton problem or to the rear side of lamella (see Fig. \ref{fig_configuration}). Keeping the Bretherton notation, this case is called the front meniscus, denoted by $f$ in the following.  
Conversely, $U<0$ corresponds to a plate plunging into a bath, to the rear end of the Bretherton's bubble, to a drop spreading on a wet plate or to the front side of a lamella, and is called the rear meniscus, denoted by $r$. 

This problem is governed by the Stokes equation, in the lubrication regime. We denote the pressure in the liquid phase $p(x)$, the velocity in the $x$ direction $v(x,y)$ and the film thickness $h(x)$. The fluid viscosity is  $\eta$ and the surface tension  $\gamma$.  
The equation of motion can then be written as: 
\be
 \drp{^2v}{y^2}=\frac{1}{\eta} \drp{p}{x} = -\frac{\gamma}{\eta} \drp{^3h}{x^3} \; .
\label{stokeslub}
\ee
Using $v(x,0)=U$ this leads to 
\be
v(x,y) = - \frac{\gamma}{ \eta} \drp{^3h}{x^3}\frac{y^2}{2} +  y \drp{v}{y}(x,0) + U \, . 
\label{vitesse1}
\ee

The continuity of tangential stress $\sigma^{int}$ provides the boundary condition at the interface.
 \be
\drp{\sigma^{int}}{x}=\eta \drp{v}{y}(x,h(x)) \; .
\label{eq_tang_stress}
\ee

The exact expression for $\sigma^{int}$ depends on the physico-chemical properties of the solution.
This interfacial stress is denoted by $\gamma$ in eq. \ref{stokeslub}, which implicitly assumes that, at dominant order in $Ca$,  $\gamma =\gamma_{eq}$,   the equilibrium surface tension. The explicit notation 
$\gamma_{eq}$ and $\sigma^{int}$ will be used to discuss the small difference between both terms. Elsewhere we will keep the generic notation $\gamma$. 
Without surfactant $\sigma^{int}= \gamma_{eq}$ is exactly verified, and the relation (\ref{eq_tang_stress}) imposes $\drp{v}{y}=0$ at the interface. This case is the stress free case denoted $(sf)$ in the following. 
In the opposite limit  of incompressible interface, the boundary condition is div$^{2D} \vec{u}=0$, with  div$^{2D}$ the divergence in the plane tangent to the interface and $\vec{u}$ the interfacial velocity. Consequently, small variations of the interfacial stress occur along the interface, discussed in section  \ref{limit_model}.
In our geometry, because the flow is quasi-parallel and invariant in the $z$ direction,  div$^{2D} \vec{u}$ simplifies into $dv(x,h(x))/dx $. The incompressibility of the interface thus implies  that $v(x,h(x))$ is constant. This uniform velocity $v(h)$ is imposed by an external condition, in the static meniscus or in the wetting film and may {\it a priori} take any value, depending on the whole experimental situation. In this paper, two particular values are considered.
The case $v(h)=v(0)=U$ will be called the rolling case $(rol)$: in the region of interest, close to the wall and to the meniscus, the interface makes a rotation around the bubble in the bubble frame and has no relative motion with respect to the wall. The case $v(h)=0$ is the sliding case $(sli)$: the interface is locally immobile in the bubble frame and slides on the moving wall.

In order to take into account these three different boundary conditions ($sf$, $sli$ and $rol$) within the same formalism, we use a general velocity expression that depends  on the two parameters $\lambda$ and $\mu$, as summarized below:

\be
v = \frac{\gamma}{ \eta} \drp{^3h}{x^3} ( \lambda hy- y^2/2) +U (1- \mu y/h) \, . 
\label{vitesse}
\ee

Finally, we impose the condition that the wetting film is flat far from the meniscus, with a thickness $\hi$. The flux (per unit length in the z direction) in the flat region is  $Q= U\hi(1-\mu/2)$ and flux conservation leads to the central relation
\be
\label{eq_motion}
\pm 3 \beta Ca (\hi-h) =  \drp{^3h}{x^3} h^3   
\ee
%beta^{sf}=1 ; beta^{slid}=2 ;   beta^{rol}=4
with $\beta = (2-\mu)/(3 \lambda-1)$ and $Ca=\eta |U|/\gamma$. By convention, we use the upper sign (here $+$) for the front meniscus and the lower sign (here $-$) for the rear meniscus. 

For the different boundary conditions we obtain:
\begin{itemize}
\item Stress free case ($sf$): $\lambda=1$, $\mu=0$, $\beta= 1$ 
\item Sliding case ($sli$):  $\lambda=1/2$, $\mu =1$, $\beta= 2$
\item Rolling case ($rol$):  $\lambda=1/2$, $\mu =0$, $\beta= 4$
\end{itemize}

\begin{figure}[h]
\centerline{\includegraphics[width=6.5cm]{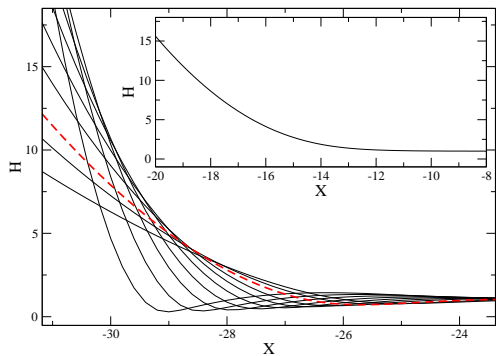} }
\caption{
Family of rear meniscus profiles, obtained by solving eq. (\ref{systeme}). The  boundary conditions at $X=0$ are derived from eq. \ref{condlim_rear} using  $\e_0 = 10^{-6}$ and different values of $\Phi$. The dashed line (red online) is the $r^*$ solution, having the same asymptotic curvature as the front meniscus. Inset: solution of eq. (\ref{systeme})  for the front meniscus. }
\label{fig_film_profile}
\end{figure}

We now define the dimensionless quantities
$h=\hi H$,  $x= \hi X/(3 \beta Ca)^{1/3}$,
and rescale  equation (\ref{eq_motion}) into 
\be
H'''H^3 = \pm (1-H) 
\label{systeme}
\ee
The boundary conditions are $H(\infty) =1$, $H'(\infty) =0$ and $H''(\infty) =0$. 
The problem is finally closed with the matching condition  $ h''(-\infty) = 1/r_m$, with $r_m$ the static meniscus radius of curvature. This condition is equivalent to 
\be
\hi = r_m (3 \beta Ca)^{2/3} H''(-\infty) \, . 
\label{hi}
\ee
A numerical resolution requires the determination  of boundary conditions at finite distances, and these are obtained from the linearized problem. 
At large $X$, $H$ is close to 1 and the problem can be linearized using a  small parameter $\e(X)= H(X)-1$. The governing equation (\ref{systeme}) becomes 
$\e'''=\mp \e$, which has  solutions of the form $\e = e^{KX}$, with $K^3=\mp 1$. For the front meniscus $K=-1$ is the only root with a negative real part and the solution $H(X)= 1+\e_0 e^{-X}$ can be used to determine the values $H(0)$, $H'(0)$ and $H''(0)$ needed to solve eq. \ref{systeme} numerically.  Changing the arbitrary value of  $\e_0$ only translates the solution along the $X$ direction. The unique   solution $H^f(X)$ is shown on Fig. \ref{fig_film_profile} (inset). 
The numerical value of  $H''_f(-\infty)$ is $H''_{f,out}=0.6430$, and, from eq. \ref{hi}, we recover the classical results \cite{shen02}
 \be
\hi^{sf}= 1.34 \, r_m Ca^{2/3} \quad;\quad   \hi^{sli}=2^{2/3}\hi^{sf}\quad;\quad  \hi^{rol}=2^{4/3} \hi^{sf} \, .
\label{film_thickness}
  \ee
In the case of a plate pulled out of a liquid, the result is usually expressed as a function of the capillary length $l_c= \sqrt{\gamma/(\rho g)}$, with $\rho$ the solution density and $g$ the gravity. Due to the hydrostatic pressure, the meniscus curvature changes with the height and can be shown to be $r_m = l_c/\sqrt{2}$ at the top of the meniscus. The film thickness is thus, in the stress free case,  $\hi=1.34 l_c Ca^{2/3}/\sqrt{2}  = 0.94\,  l_c \, Ca^{2/3}$.

\vspace{0.5cm}

For the rear meniscus, the linearized equation $\e'''= \e$ admits two roots with a negative real part $K=-1/2 \pm i\sqrt{3}/2$ . 
The general solution of the linearized equation is thus
\be
\e= \e_0 e^{-X/2} \cos(\sqrt{3}X/2+ \Phi) \; .
\label{condlim_rear}
\ee
Different values of $\e_0$ just translate the solution, but different values of $\Phi$ lead to different solutions, with $H''$ converging to a large range of possible values (positive or negative)  when $X \to -\infty$  \cite{giavedoni99}.  The  family of  rear meniscus profiles are represented on Fig. \ref{fig_film_profile}. 
This states that, for a given meniscus radius of curvature, we can find a steady solution for any film thickness initially present on the wall.
In the following discussion, the profile with $H''(-\infty) = 0.643$, as for the front meniscus, will be denoted by the subscript $r^*$. It plays an important  role, as it is the shape obtained when the rear meniscus moves on the wetting film deposited on the wall by a front meniscus of same radius, moving at the same velocity.

\subsection{Asymptotic parabolic shapes}
\label{parabole}

\begin{figure}[h]
\centerline{\includegraphics[width=7cm]{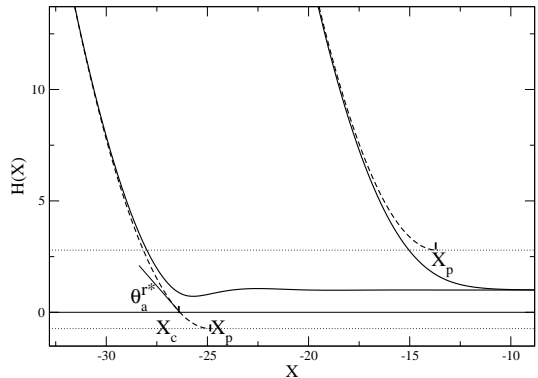} }
\caption{Asymptotic parabolic shapes of the meniscus.  The solid lines are the numerical solutions of eq. \ref{systeme} for the rear meniscus $r^*$ (left) and for the front meniscus (right). The dashed lines are the best parabolic  fits. The horizontal dotted lines show the height of the parabola minimum $H_p^{r^*}=-0.82$ and   $H_p^f=2.88$, for the rear and front cases, at the position $X_p$. The apparent contact point of the rear meniscus is $X_c$ and the apparent contact angle at this point is $\theta^{r^*}_a$. }
\label{fig_asympt_profile}
\end{figure}

The asymptotic expression for the film thickness for $X\to - \infty$ is a parabolic profile $H_{out}(X) = H''(-\infty) (X-X_p)^2/2  + H_p$, with $X_p$ the position of the minimum of the parabola and $H_p= H_{out}(X_p)$ \cite{duffy97}. Both $X_p$ and $H_p$  need to be determined numerically. The best numerical fit, plotted on Fig. \ref{fig_asympt_profile},  have been obtained  with $H^f_p= 2.88$ for the front meniscus with  $X$ in the range $[-400; -150]$. 
A slightly smaller value, $H^{f, Br}_p= 2.79$ is reported in \cite{bretherton61} and is found when the fit is performed  for $X$ in the range $[-90; -40]$.

For the rear meniscus, we found $H^{r^*}_p= -0.82$.
Its asymptotic profile thus intersects the wall at the contact  point $X_c$ shown in Fig. \ref{fig_asympt_profile}.  The apparent contact angle $\theta^{r}_a$, also shown in  Fig. \ref{fig_asympt_profile}, has been computed for any initial wetting film thickness in \cite{chebbi03}. 
It can be deduced from the parabola equation: $H'_{r,out}(X_c)= \sqrt{-2\, H^{r}_p \, H''_{r,out}}$ and from the relation $(3 \beta Ca)^{1/3} H'_{r,out}(X_c)= \tan \theta^{r}_a \sim \theta^{r}_a$. For the three boundary conditions considered, we get, for the $r^*$ solution:

\be
\theta^{sf,r^*}_a= 1.4 \,  Ca^{1/3}  \quad ; \quad \theta^{sli,r^*}_a= 1.7 \, Ca^{1/3} \quad ; \quad \theta^{rol,r^*}_a= 2.2\,  Ca^{1/3}
\label{angleapp}
\ee

These relations are very similar to Tanner's law, which  describes the spreading of a droplet on a solid wall in the perfect wetting case \cite{tanner79,degennes85,bico01,cormier12}. However, for a spreading droplet, the profile far from the wall is not parabolic, but rather close to a straight wedge. In that case, the apparent contact angle  involves a  logarithmic term.
Indeed, imposing the boundary conditions  $\hi= 0$ when $x\to \infty$ and $h''(x)\to 0$ when $x\to -\infty$ in eq. (\ref{eq_motion}) leads to \cite{tanner79,degennes85}
\be
\theta^{drop}_a(x)= [9 \ln(x/x_{min})]^{1/3} Ca^{1/3} \; .
\ee
The micro-scale cut off is obtained by matching the film profile to a precursor film initially present  on the wall and governed by Van der Waals forces : $x_{min} = 2.5 \bar{A}/(6 \pi \gamma) 3^{-1/6} Ca^{-2/3}$, with $\bar{A}$ the Hamaker constant \cite{degennes85}. 

 \vspace*{0.5cm}

\subsection{Comparison with some experimental data}

\subsubsection{Wetting film thickness in the LLD geometry}

The wetting film thickness as a function of the plate  velocity  has been extensively studied for the front meniscus in the LLD geometry, with fibers or plates pulled out of pure liquids or surfactant solutions. 
The stress free result  $\hi^{sf}$ is obtained for pure liquids.  For surfactant solutions, the whole range of thicknesses between $\hi^{sf}$ and $\hi^{rol}$ are observed, depending on the surfactant concentration, the imposed velocity and, for the fiber case,  the fiber radius  \cite{quere99, scheid10}.
In LLD geometries, surfactant solutions with high surface modulus  lead to a wetting  film thickness in good agreement with  the rolling case.  
While the solid is pulled out  of the liquid reservoir, the total interfacial area increases: the new area is produced at the free interface of the liquid reservoir, in agreement with the rolling assumption. This is confirmed by  recent velocity field measurements in the reservoir \cite{mayer12}. \\

The sliding case implies that the whole wetting film is sheared. It  is thus incompatible with a very long wetting film  (see section \ref{validity_size} for a quantitative discussion). It is therefore not observed in the LLD geometry, but can be obtained for bubbles, as discussed below. 

\subsubsection{Film profile for bubbles with incompressible interfaces}
\label{section_profile_denkov}

\begin{figure}[h]
\centerline{\includegraphics[width=8cm]{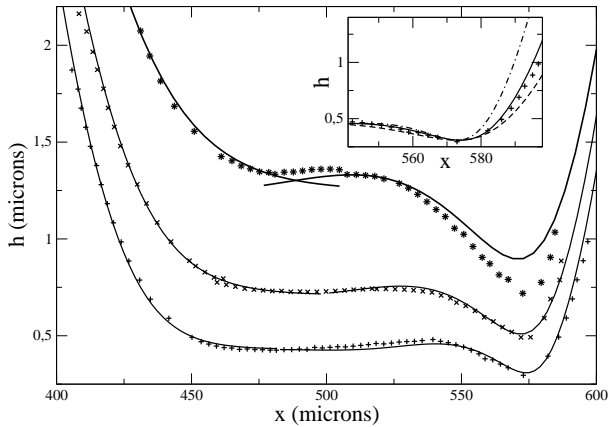} }
\caption{Wetting film profiles for a 2D bubble in contact with a wall moving to the right, for incompressible interfaces. $+$,  $\times$ and $\bullet$: experimental data from  Fig 12a in \cite{denkov06},   for $Ca= 1.15\, 10^{-4}$, $2.3\,  10^{-4}$ and $4.6\, 10^{-4}$, respectively. Full lines:  solutions of eq. (\ref{systeme}) (shown in Fig. \ref{fig_film_profile}) rescaled  using $\beta=2$ (sliding case). Insert: Zoom on the rear meniscus of the first profile. Dashed, full and dash-dotted  lines:  solutions of eq. (\ref{systeme}) rescaled  using $\beta=1$ (stress free case, red online), $\beta=2$ (sliding case) and $\beta=4$ (rolling case, blue online) respectively.}
\label{fig_profileexp}
\end{figure}

The film profiles shown in Fig. \ref{fig_profileexp} were measured by Denkov {\it et al.} using interferometry, for a solution of very high dilatational surface modulus \cite{denkov06}.  We compare
 these experimental results  with the rear and front solutions of eq. (\ref{systeme}). As the meniscus curvature $r_m$ is the same on both sides we use the $r^*$ solution for the rear meniscus. The coordinates   $x=\hi X/(3 \beta Ca)^{1/3}+ x_0$ and $h= \hi H$ are deduced from the numerical solution $H(X)$ using  the experimental  capillary number $Ca$,  the film thickness $\hi$ (measured in the middle of the film) and a translation parameter  $x_0$  that is independently fitted for the rear and front parts of the profile. The three possible values of $\beta$ have been tested.
For the two slower cases,  the rear and front solutions nicely overlap on the central region  and a very  good fit is obtained using $\beta=2$ (sliding case). The profiles obtained for $\beta=1$ (stress free) and $\beta=4$ (rolling) are also plotted on Fig. \ref{fig_profileexp}(inset) for one meniscus and show a significatively worse agreement with the experimental profile. It thus confirms that the foaming solution used, containing insoluble surfactants, leads to a sliding regime in this geometry, as already assumed in \cite{denkov06}. 

At larger velocities, the contact area between the bubble and the wall, {\it i.e.} the area of the central flat film, decreases and eventually disappears. The two solutions
$H^{r^*}(X)$ and $H^f(X)$ are both determined by assuming a flat film far from the meniscus and are thus not valid anymore. In that regime, the whole hydrodynamical problem, involving the two meniscus at the same time, should be solved. 

\vspace*{0.5cm}

\subsubsection{Flux around a bubble for pure water}
\begin{figure}[h]
\centerline{\includegraphics[width=7cm]{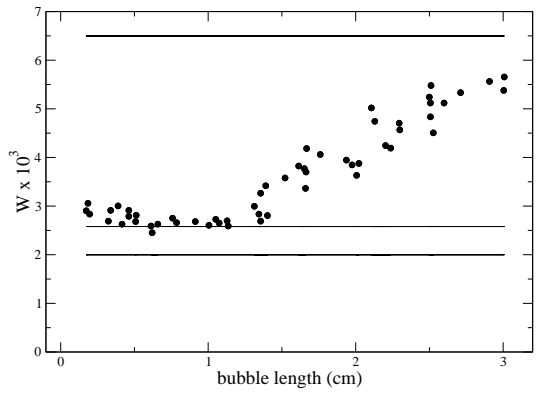} }
\caption{Fractional speed difference $W$ of the liquid phase around an isolated bubble in a tube. $\bullet$ : experimental data from Fig. 4 \cite{schwartz86}, with $Ca=3\,10^{-5}$ and $r_t=0.5$mm. The solid lines are the predictions, from top to bottom,  for the rolling case, the stress-free case and the sliding case. }
\label{fig_schwartz}
\end{figure}

An indirect measurement of the wetting film thickness has also been obtained by Schwartz, Princen and Kiss \cite{schwartz86} for the case of a bubble in pure water. They pushed a single bubble along  a tube of radius $r_t$ and measured the solution flux in the bubble frame $Q_{tot}$. They report  a fractional speed difference  $W$ defined by $W=Q_{tot}/(\pi r_t^2 U)$.  From eq. \ref{vitesse}, we predict $Q_{tot}=2 \pi r_t U \hi (1-\mu/2)$ and  $W= 2 \hi(1-\mu/2) /r_t$. Finally we get  $W^{sf}= 2.68 Ca^{2/3}$,  $W^{sli}= 2^{-1/3}W^{sf}$ and $W^{rol}= 2^{4/3}W^{sf}$, that we compare with the experimental data in Fig. \ref{fig_schwartz}.
The flux measured for short bubbles is in nice agreement with the stress free prediction, as expected for pure water. However, for long bubbles (more than 1.5 cm  for a 1mm inner diameter tube) the flux increases progressively, which is in  contradiction with a transition toward a sliding case, for which a smaller flux is expected. In that case, the experimental observation is instead compatible with a transition toward a rolling case, as observed in LLD geometry \cite{park92}. 

It is not clear why the regime is modified with the bubble length in this  case and, more generally,  what determines the sliding or rolling regime in the limit of incompressible interfaces.
Note that for a bubble in a cylindrical tube, the flow is necessary axisymmetric.  The rolling case is thus not compatible with the incompressibility of the interface at the global scale, as some interface needs to be created at the front and suppressed at the rear of the bubble.

\vspace*{0.5cm}

\section{Viscous forces in a meniscus}
\label{section_force}

\subsection{Theoretical predictions}
\label{theorie_fv}
 
\begin{figure}[h]
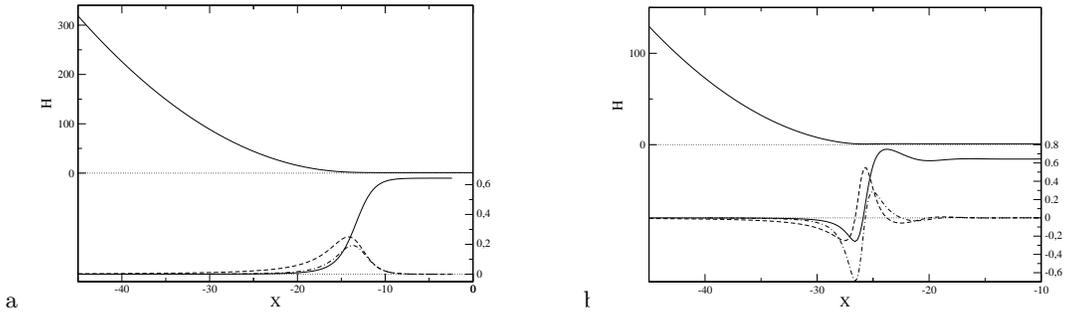

\centerline{a \; 
\includegraphics[width=6cm]{fig7a.eps} \hspace*{1cm} b\; 
\includegraphics[width=6cm]{fig7b.eps} }
\caption{Normal and tangential force distribution in the front and $r^*$ meniscus, in dimensionless units. (a) Upper graph: Front meniscus profile. Lower graph: solid line:   pressure $P=0.643-H''(X)$; dashed line: meniscus contribution to the viscous stress $(H-1)/H^2$; dot-dashed line: velocity divergence $-H'/H^2 $. (b) Same functions for the rear meniscus $r^*$. Full line: $P=0.643-H''(X)$; dashed line: $ (1-H)/H^2$; dot-dashed line: $H'/H^2 $.}
\label{fig_local_force}
\end{figure}

The viscous force exerted by the solid wall on the liquid (per unit length of meniscus in the $z$ direction) can be deduced from the velocity expression eq. \ref{vitesse}.
\be
f^{f/r} =  - \eta \int_{meniscus} \drp{v}{y}(x,0) dx =
 -\eta \int_{-\infty}^{\infty} \lambda \frac{\gamma}{\eta}\drp{^3h}{x^3} h  dx  + \mu U \eta \int_{-\infty}^{\ell}\frac{1} {h}  dx \, .
\ee
The first term converges at $+\infty$ and $-\infty$ and corresponds to a viscous stress localized in a small domain close to the apparent contact point between the meniscus and the wetting film, as shown in dimensionless form on Fig. \ref{fig_local_force}. This domain is usually called the dynamical meniscus, and its  extension $\ell_d$ scales with  $\hi/(3 \beta Ca)^{1/3}$, {\it i.e.} as $r_m Ca^{1/3}$, with a prefactor discussed in section \ref{validity_size}. 
 
 The second term  diverges at $+\infty$ and only appears for the sliding case ($\mu =1$) for which the whole wetting film is sheared. It thus  imposes  a cut-off length $\ell$ corresponding to the actual length of the wetting film. 
Expressed in term of the universal function $H$ the force (per unit length) is given by:
\be
f^{f/r} = \gamma (3 \beta Ca)^{2/3} \left[ \lambda F^{f/r}  +   \frac{\mu }{3\beta} G^{f/r}(L) \right]
\label{f_v_gene}
\ee
with $F^{f/r}= \mp\int_{-\infty}^{\infty}(1-H)/H^2$,  $G^{f/r}(L)=\pm \int_{-\infty}^{L}1/H$ and $L = (\ell/r_m) (3 \beta Ca)^{-1/3}/H''_{f,out}$. As $L$ depends on $Ca$, the last term introduces some additional $Ca$ dependence into the viscous force.

\vspace*{0.5cm}

\vspace*{0.5cm}
\begin{figure}[h]
\centerline{\includegraphics[width=6.5cm]{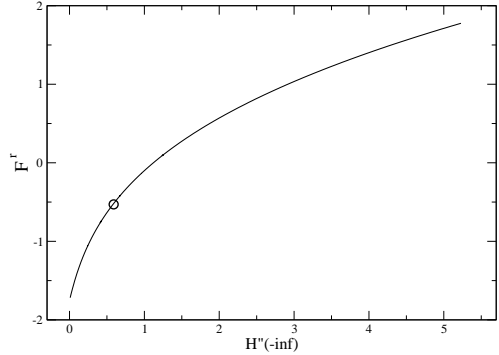} }
\caption{$F^r= \int_{-\infty}^\infty (1-H)/H^2 dX$ as a function of the asymptotic curvature $H''(-\infty)$. $F^r$ is the dynamical meniscus contribution to the viscous force per unit length exerted on the rear meniscus. It is related to the corresponding  physical quantity by the eq. \ref{f_v_gene}. The circle represents the $r^*$ solution.  }
\label{fig_Fvar_de_Hsec}
\end{figure}

In the front meniscus, $H>1$ everywhere, implying that the viscous force exerted by the wall  is positive, as expected for $U>0$. The numerical integration leads to $F^{f}= 1.85$.
In comparison, the rear meniscus is oscillating around $H=1$ and, as underlined in \cite{bretherton61}, the sign of the force depends on the value of $H''(-\infty)$. We discuss this point in section \ref{pressure_singleB}.
The value of $F^r$ is plotted as a function of $H''(-\infty)$ in Fig. \ref{fig_Fvar_de_Hsec}.
For  the $r^*$ profile,  we find $F^{r*}= -0.53$ and thus,  as for the front meniscus,  a viscous force exerted by the wall in the direction of its own motion. 
The vanishing force corresponds to $H''(-\infty)= 1.1$, which implies $\hi= 1.1 \, r_m (3 \beta Ca)^{2/3}$.

The wetting film contribution $G^{f/r}(L)$ tends to an affine function of $L$ at large $L$, because $H(X)$ becomes constant. The coefficients of the affine law depend on the precise definition of  $L$. In this case, we choose $L$ as the distance between $X_p$, the minimum of the asymptotic parabolic profile (see Fig. \ref{fig_asympt_profile}), and the end of the wetting film. With this definition, we get, for long wetting films:
\be
G^f= L +  1.27   \mbox{  and  }   G^{r*}= - L - 4.0
\label{G}
\ee
The agreement with the exact numerical solution is within a few percent for $L>2$.

Finally,  the viscous force per unit length for each investigated case is, in physical units:

\begin{align}
& f_{sf}^{f}= 3.84 \gamma Ca^{2/3} \hspace*{1cm}  f_{sf}^{r^*}=  -1.1 \,  \gamma Ca^{2/3} \label{fsf}\\
& f_{sli}^{f} = 3.75 \gamma Ca^{2/3} + 0.47 \gamma Ca^{1/3}  \frac{\ell}{r_m} \hspace*{1cm} f_{sli}^{r^*} = -3.07 \gamma Ca^{2/3} - 0.47 \gamma Ca^{1/3}  \frac{\ell}{r_m}  \label{fsli}\\
& f_{rol}^{f} = 4.84  \gamma Ca^{2/3} \hspace*{1cm}  f_{rol}^{r^*}=  -1.38 \,  \gamma Ca^{2/3} \label{frol}
\end{align}

Given the orientation conventions used in the paper, for a wall moving in the positive direction, the force per unit length exerted on a bubble or on a lamella by the wall is given by $f=f^{f} - f^{r^*}$. We define the parameters $\xi=(4.94 ; 6.82 ; 6.22)$ and $\zeta=(0 ;  0.94 ; 0)$ respectively for the stress free, sliding and rolling cases, so that 
\begin{equation}
f= \xi \gamma Ca^{2/3} + \zeta \gamma Ca^{1/3}  \frac{\ell}{r_m} \, . 
\label{force_tot}
\end{equation}

The viscous force exerted on a meniscus  thus exhibits  a power law in relation to the velocity, with an exponent of 2/3 for the stress free case and for the rolling case. However, for the sliding case, the force  is a combination of two power laws of exponents 1/3 and 2/3. 
For a typical capillary number $Ca= 10^{-3}$, both terms are of the same order of magnitude if the meniscus size $r_m$ is 10 times smaller than the bubble size $\ell$. 
This latter  result provides some theoretical clarification to the seminal work by Denkov {\it et al.} which first shed light on the specific behavior of foams with incompressible interfaces. The approximate resolution of eq. \ref{eq_motion} for the sliding case developed in  \cite{denkov05} relies on an approximated shape for the wetting film and on the assumption that its  thickness is governed by a minimum dissipation principle. It leads to the prediction that  the contribution of the sheared wetting film to the friction force scales as $\gamma (\ell/r_m)^{1/2} Ca^{1/2}$. A 1/2 power law is found for deformable objects with elastic properties \cite{meeker04}, but, as shown in this paper, it does not apply to the friction between  bubbles and  a solid, even in the incompressible interface limit.

The two  power laws appearing in eq. (\ref{force_tot}) can be derived from simple scaling arguments. The term localized  in the dynamical meniscus is built from  a velocity gradient $U/\hi$, integrated along the length $\ell_d \sim  \hi Ca^{-1/3}$. This leads to a force per unit length varying as $\eta U Ca^{-1/3} = \gamma Ca^{2/3}$. The term obtained from the shearing of the whole wetting film of length $\ell$ scales as $\eta \ell U/\hi \sim \gamma (\ell/r_m) Ca^{1/3}$.

Finally, the viscous forces  obtained in the rolling and stress free cases are simply related to each other by the relation $f_{rol}^{f/r} = 2^{1/3} f_{sf}^{f/r}$. This  is not the same rescaling as for the film thickness $\hi$, found to be  $\hi^{rol}=2^{4/3} \hi^{sf}$.

 In the frame of this model, in these two cases, the viscous force per unit length of meniscus  does not depend on the meniscus radius $r_m$  \cite{ratulowski89}.

\subsection{Validity range of the models}
\subsubsection{Velocity range}
 \label{velocity_range}
At very small capillary numbers, or at a very small meniscus radius of curvature, the wetting film thickness predicted by eq. \ref{film_thickness} becomes so small that the disjoining pressure can not be neglected anymore. Teletzke {\it et al.} have shown in \cite{teletzke88}, that the wetting film  and meniscus profiles become independent of the capillary number for very small $Ca$, 
when they  have the same shape as at equilibrium, as observed in \cite{quere89}. We thus expect a viscous force that increases linearly with $Ca$, as is the case for non deformable interfaces in Stokes regime.

At large velocity, inertia becomes important. Qu\'er\'e {\it et al.} \cite{quere98, aussillous00} measured the thickness of the liquid film deposited on a fiber of radius $r_f$ pulled at constant velocity out of a liquid bath. They found a deviation from the theory for $Ca > 10^{-2}$. At larger velocity, they predict that the film thickness and the dynamical meniscus extension scale as  $\hi \sim r_f Ca^{2/3}/(1-We)$ and $\ell_d \sim r_f Ca^{1/3}/(1-We)$, where $We= \rho U^2 r_f/\gamma$ is the Weber number.  
The viscous force involves (in the stress free and rolling cases) the ratio $\ell_d/h$. As these two length scales have the same inertial correction, we thus expect that the viscous force departs from the theory at larger capillary numbers than the film thickness. 
Experimentally, the $Ca^{2/3}$ scaling has been observed for the viscous force up to $We \sim 10$ for films in cylindrical tubes \cite{dollet10}.

\subsubsection{Interfacial rheology}
 \label{limit_model}

The assumption of incompressible interface or of vanishing tangential stress are almost never perfectly verified and we estimate here the experimental conditions in which these conditions are achieved within the desired precision. 

The interfacial extension rate is given by
\be 
\frac{d v(x,h(x))}{d x}=\left(\lambda-\frac{1}{2}\right) \frac{\gamma}{\eta} \frac{\partial}{\partial x} \left( h^2\drp{^3h}{x^3} \right) = \mp  \left(\lambda-\frac{1}{2}\right) \frac{\gamma}{ \eta r_m} \frac{(3 \beta Ca)^{2/3}}{ H''_{f,out}}
 \frac{H'}{H^2} \, . 
\label{divv}
\ee
For the incompressible case, $\lambda=1/2$ and the extension is thus consistently 0. However, the extension  reaches large values for the stress free case. The functions $\mp H'/H^2$ are plotted on Fig. \ref{fig_local_force}  and have a maximum at 0.19 for the front meniscus and a minimum at -0.7 for the rear meniscus. For $U=1$cm/s, $\eta= 10^{-3}$ Pa.s, $\gamma = 30 \, 10^{-3}$ N/m ({\it i.e.} $Ca= 0.3 \,10^{-3}$)    and $r_m = 2\, 10^{-4}$m, the maximal extension or compression rate is thus of the order of 400 s$^{-1}$.
The stress free model is thus applicable  if no noticeable surface tension variations are observed for this  extension rate.
Considering an interface for which the interfacial stress variations are dominated by the  interfacial dilatational viscosity $\eta_s$, and a desired precision of $(\sigma^{int}-\gamma_{eq})/\gamma_{eq} \ll \epsilon$, we obtain the validity  criteria for the stress free model as
\be
 \frac{\eta_s}{\eta r_m} Ca^{2/3} \ll \epsilon \; . 
\ee
With the previous numerical values, the  relative variation in interfacial stress is $1\%$ for $\eta_s= 0.7\, 10^{-6}$ Pa$\cdot$m$\cdot$s.

% 400 eta_s / 30 10-3  = 10-2
%eta_s = 30 10-5/400 = 0.7 10-6

A similar approach can be used to determine the applicability range of the incompressible model. 
For the rolling or sliding cases, the problem is solved without specifying the tangential stress  at the interface $\sigma^{int}$. 
However, $\sigma^{int}$ can be determined {\it a posteriori} from the tangential stress continuity (\ref{eq_tang_stress}).
Substituting  the velocity field defined in  eq. \ref{vitesse} in the relation \ref{eq_tang_stress} leads to 
\be
\drp{\sigma^{int}}{x}=\gamma \left(- h\drp{^3h}{x^3} (1 - \lambda) - \mu \frac{\eta U}{\gamma} \frac{1}{h}\right ) = \mp \frac{\gamma}{r_m} \frac{(3 \beta Ca)^{1/3}}{H''_{f,out}} \left( (1- \lambda)\frac{1-H}{H^2} + \frac{\mu}{3 \beta H} \right)
\ee
As expected,  $\sigma^{int} $ has a constant value for the stress free case ($\lambda=1$, $\mu=0$). When $\lambda=1/2$ (incompressible case), variations of order $Ca^{2/3}$ 
occur. Indeed, by integration over $x$ between $-\infty$ and $\ell$, we get:
\be
\delta \sigma^{int}= \sigma^{int}_{\ell}- \sigma^{int}_{-\infty} = \gamma (3 \beta Ca)^{2/3} \left[ (1-\lambda) F^{f/r} - \frac{\mu}{3 \beta} G^{f/r}(L)\right]
\label{dgammma}
\ee

This incompressible model will be suitable for interfaces able to produce such interfacial stress variations without noticeable surface extension. In the rolling case ($\mu=0$), the interfacial stress difference  $\delta \sigma^{int}$ equals the viscous force per unit length exerted by the wall.
This is a direct consequence of the fact that the velocity field is a parabola centered in the middle of the film.
For the same parameter values as above, we get $\delta \sigma^{int}= 0.75$mN/m. In the case of an interface dominated by elasticity, $\delta \sigma^{int}  =E \delta A/A$, and for the interface extension to be less than $1\%$, the dilatational modulus of the interface, $E$, must be larger than 75 mN/m. More generally, in the rolling case,  $\delta A/A \ll \epsilon$ if
\be
\frac{\gamma}{E}Ca^{2/3} \ll \epsilon \, , 
\ee
which is thus the validity criteria of the incompressible model, at the desired precision $\epsilon$.

The sliding case involves the wetting film length and  will be discussed in the next section.

% 
% $U = 10^{-2} m/s$ \\
% $\eta= 10^{-3} Pa s$ \\
% $\gamma = 30 10^{-3} N/m $\\
% $3 Ca= 10^{-3}$\\
% $r_m= 2 10^{-4} m$\\
% $\hi =  10^{-6} m $\\

\subsubsection{Meniscus size and length of the wetting film}
\label{validity_size}
The dynamical meniscus is  defined as the place where the viscous friction takes place (in the rolling or stress free cases). Figure \ref{fig_local_force} allows us to determine the extension $\ell_d$ along $x$, of this dynamical meniscus.  
Imposing the restriction that only $5\%$ of the viscous force is localized on each side of the dynamical meniscus, we find that the dynamical meniscus, in dimensionless form, corresponds to the $X-X_p$ range $[-32 ; 3]$  for the front meniscus and $[-75; 3]$ for the $r^*$ one. So $\ell_d^f= 35 \hi /(3 \beta Ca)^{2/3} $ and $\ell_d^r= 78 \hi /(3 \beta Ca)^{2/3}$.
 In physical units, for a typical  capillary number of $Ca= 0.3\, 10^{-3}$, this leads to an extension of the dynamical meniscus of the order of $ 0.2\,r_m$, from the apparent contact point toward the wetting film. In a bubble geometry, as in Fig. \ref{fig_profileexp}, the two menisci are thus independent if
$\ell > 0.2 \, r_m$ at $Ca= 0.3\, 10^{-3}$.
This condition becomes less restrictive for smaller Ca. 

On the meniscus side ($X-X_p<0$), we obtain an extension which is of the order of the meniscus itself for $Ca= 0.3\, 10^{-3}$. 
The fluid flow in the meniscus is thus probably not always negligible. This may be why the viscous force measured for lamella is 
higher than the force found for isolated bubbles \cite{terriac06,raufaste09,dollet10}.

In the sliding case, not only the lower bound for the wetting film length should be considered, but also an upper bound. 
In this specific  case, the interfacial stress variation (\ref{dgammma}) increases rapidly with $\ell$ and is dominated by the second term for long bubbles. 
The model derived from eq. \ref{stokeslub} is not valid for large variations of the interfacial stress. Indeed, in such cases, the Laplace pressure derivative should involve the derivative of the curvature,  but also the derivative of $\gamma$. 
However, an order of magnitude of the maximal length compatible with the sliding regime is estimated by assuming that 
 $\delta \sigma^{int}$ is of the order of $\gamma_{eq}$. Injecting this condition in eq. \ref{dgammma} leads to:
\be
\ell_{max} \approx 2 r_m Ca^{-1/3}
\label{lmax}
\ee
For $Ca= 0.3 \, 10^{-3}$ this leads to $\ell_{max}\sim 30 r_m$.
This theoretical limit has been observed experimentally in \cite{cantat12}.

\section{Pressure drop for an isolated bubble or a lamella}
\label{section_pressure}

\subsection{Force balance}

\subsubsection{Single bubble}
\label{pressure_singleB}
\begin{figure}
\centerline{
\includegraphics[width=7cm]{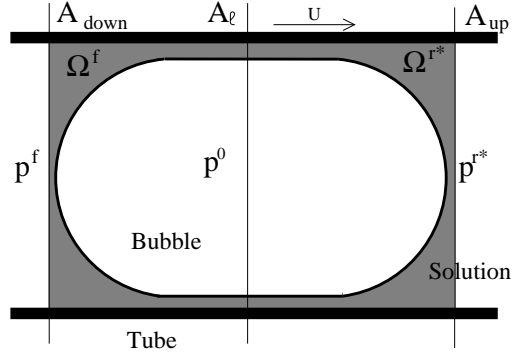} 
}
\caption{Sketch of a single bubble in a tube, with the two subsystems $\Omega^f$ and $\Omega^{r^*}$ used for the force balance. The two subsystems are defined by the liquid bounded by the tube wall, the median plane $A_\ell$  and, respectively, the plane $A_{down}$ and $A_{up}$.  }
\label{fig_force_balance}
\end{figure}

We now consider a single bubble in a capillary of constant section ${\cal S}$  and of perimeter ${\cal P}$. We assume that the bubble is long enough to create a thin wetting film between the bubble and the wall (see Fig. \ref{fig_force_balance}). The problem is not invariant in the $z$ direction anymore. However the results of the previous section concern the flow in a small region between the meniscus and the wetting film of  extension $r_m\, Ca^{1/3}$. These results thus remain  valid as long as the tube dimensions  are much larger than $r_m\, Ca^{1/3}$. 
Some corrections should nevertheless be considered if the tube has an angular cross-section \cite{wong95a, wong95b, hazel02,cubaud04}.

In Fig. \ref{fig_force_balance}, we define three planes  perpendicular to the tube section: $A_{up}$ and $A_{down}$ are close to the rear and front meniscus, but outside the dynamical meniscus, and
$A_\ell$ is the plane  in the middle of the bubble, situated  at a distance $\ell$ from each effective contact point between the meniscus and the wetting film.  As already discussed for the film profiles in Fig. \ref{fig_profileexp}, if the bubble is long enough, the wetting film properties close to $A_\ell$ are very close to their asymptotic values at $+\infty$, which are identical for the front and $r^*$ films. For menisci in a sliding regime, $\ell$ represents the length of wetting film associated with each meniscus (see section \ref{theorie_fv}). 

The force balance made on the open system $\Omega_f$ (resp. $\Omega_{r^*}$) constituted by the liquid between  $A_{down}$ (resp. $A_{up}$) and $A_\ell$, leads, if momentum fluxes are negligible, to the relation
\be
%{\cal P}  f^{f/r}+ \gamma(\ell){\cal P} + {\cal S} ( \langle p^{f/r} \rangle -p^0)  =0
p^{f/r} = p^0 - \frac{\cal P}{\cal S}(\gamma(\ell)+f^{f/r})
\label{p_drop}
\ee
with $p^0$ the pressure in the bubble and  $p^{f/r}$ the average pressure, respectively  in the section $A_{down}$ and  $A_{up}$. In the static case, $f^{f/r}=0$, and the relation predicts the Laplace under-pressure in the liquid phase. In the dynamical regime,  the pressure in the liquid  is smaller than the static value for the front meniscus  and higher for the rear meniscus, as $f^{f}>0$ and $f^{r}<0$ (see eqs. (\ref{fsf}-\ref{frol})). 
Note that the pressure and velocity fields have been obtained in section \ref{section_force}  by assuming a static pressure
in the meniscus. Thus, obviously, the dynamical pressure drop can  not be deduced from the pressure distribution plotted  on Fig. \ref{fig_local_force}.
 The dynamical pressure drop appears as a higher order correction that could in principle be re-injected as boundary condition for the film profile, in an iterative scheme \cite{park84}.
 
The total pressure drop for an isolated bubble is $\delta p=  p^{r^*}  -   p^f  =({\cal P}/ {\cal S}) ( f^{f} - f^{r^*}) = ({\cal P}/ {\cal S}) f $. 
For the case of a cylindrical tube of radius $r_t$ the previous relation simply becomes $\delta p = 2f/r_t$, so from eq. \ref {force_tot}:

\begin{align}
& \delta p^{sf}= 4.94 \,  Ca^{2/3} \, \frac{2 \gamma}{r_t} \label{dp_fv}\\
& \delta p^{sli}=   \left(6.82 Ca^{2/3} + 0.94 \frac{\ell}{r_m} Ca^{1/3}\right) \frac{2 \gamma}{r_t}\label{dp_fv_sli}\\
& \delta p^{rol}= 6.22 \, Ca^{2/3} \,   \frac{2 \gamma}{r_t} \label{dp_fv_rol}
\end{align}
The radius $r_t$ and $r_m$ are identical for the bubble, but differ for the lamella case, discussed below. 
 
This pressure drop, related to the presence of the meniscus, can be compared to the pressure drop induces by the Poiseuille flow in the tube, on both sides of the bubble $\delta p^{Pois}= 
8 \eta L_t U /r_t^2$, with $U$ the average liquid velocity and $L_t$ the tube length. For $r_t=1$mm and $Ca=10^{-3}$, the meniscus contribution (in the stress free case) is of the same order than the pressure drop induced by a length $L_t=12$mm  of tube. The local force induced by the meniscus thus strongly dominates the confined diphasic flows, even for relatively distant bubbles.  \\

As discussed in section \ref{theorie_fv}, the sign of the viscous force is not obvious and may vary for the rear meniscus. 
An energy balance is made below (for the stress free case), that provides a condition for this sign \cite{bretherton61}.

First, the mass balance on $\Omega^f$ or  $\Omega^r$ leads to
\be
{\cal P} \hi U = {\cal S} U'
\label{massB}
\ee
 with $U'$ the average velocity on the plane $A_{down}$ or $A_{up}$. The energy balance is 
\be
 p^{f/r}  {\cal S} U' - p^0 {\cal P} \hi U +  {\cal P} f^{f/r} U+  {\cal P} \gamma U -  {\cal P} \gamma U >0 \, . 
\ee
The last two terms are respectively the power provided by the interface and the free energy flux exiting the open system $\Omega$. 
Replacing eq. \ref{p_drop} and eq. \ref{massB} into this equation leads to the condition, at dominant order 
 
%\be-  U \frac{\hi{\cal P}}{ {\cal S}}  (f^{f/r}+ \gamma)   +   f^{f/r}U >  0 \, . \ee

\be
U f^{f/r} >  U  \gamma \frac{\hi{\cal P}}{ {\cal S}}  \, . 
\ee
If $U>0$, this leads to $f^f  > \gamma \hi{\cal P}/ {\cal S}>0$. In contrast, for $U<0$, we get $f^r < \gamma \hi{\cal P}/ {\cal S}$. The viscous force can thus, {\it a priori}, take positive or negative values. Using $\hi= H''(-\infty)r_m(3 Ca)^{2/3}$ (eq. \ref{hi}) and $f^r= \gamma (3 Ca)^{2/3} F^r$ (eq. \ref{f_v_gene}) and  taking  ${\cal P}/ {\cal S} = 1/r_m$ which is the most restrictive condition, we get 
\be
F^r <  H''(-\infty)
\ee  
which is verified by the numerical result shown on Fig. \ref{fig_Fvar_de_Hsec}.

\subsubsection{Lamella}

\begin{figure}
\centerline{
\includegraphics[width=7cm]{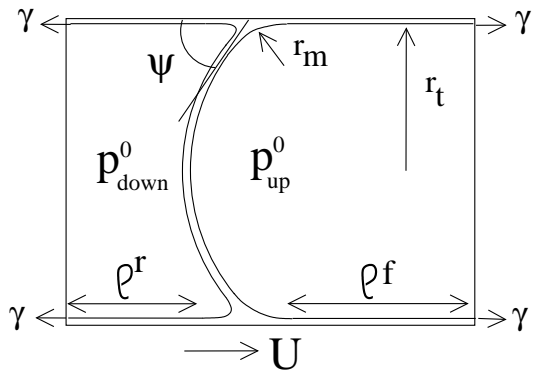} 
}
\caption{Sketch of a lamella.}
\label{fig_force_lamella}
\end{figure}

Considering the  case of a single lamella, the situation only differs in a  few details. The rear meniscus is downstream and the front meniscus upstream. A first consequence of this  is that the wetting film downstream can be of arbitrary thickness, and the viscous force corresponding to the adapted rear profile must be used (see fig. \ref{fig_Fvar_de_Hsec}). A force balance carried out  on the open system depicted on Fig. \ref{fig_force_lamella} leads to the relation  $\delta p= p_0^{up}- p_0^{down} = (2/r_t)  (f^{f} - f^{r} + \gamma^{f}(\ell^f)- \gamma^r(\ell^r))$  for a cylindrical tube.  In contrast with the bubble case, the force balances made for the rear and front menisci involve the surface tension at two distinct points and the difference $\gamma^{f}(\ell^f)- \gamma^r(\ell^r)$ is thus {\it a priori} unknown for the incompressible cases.  
However, if the external boundary conditions impose the same surface tension at both sides of the lamella, we get the same results as for  an isolated bubble, eqs. (\ref{dp_fv}-\ref{dp_fv_rol}).

For lamella, another limitation must be considered, that the meniscus is not pushed by a pressure difference, but pulled by the lamella. The force exerted by the lamella, per unit length in the $z$ direction and projected in the $x$ direction, is $2\gamma \cos \psi$, with $\psi$ the apparent angle between the lamella and the wall, defined in Fig. \ref{fig_force_lamella}. This force per unit length  can not exceed $2\gamma$, which  thus determines the maximal velocity $U_m$ at which a lamella can be pulled \cite{dollet10}:

\be
f(U_m) = 2 \gamma \, .
\label{um}
\ee

In the stress free case, this leads to the simple condition $Ca < 0.26$, or, for $\gamma = 30 \, 10^{-3}$ N/m and $\eta= 10^{-3}$Pa$\cdot$s, $U_m= 8$m/s. Above this maximal velocity, the lamella breaks. For foam transport in porous media, as used for oil recovery or soil depollution, this dynamical criteria for the lamella stability may be an important issue.

\subsection{Bretherton's pressure derivation for a single bubble}

In the specific geometry of a single bubble in a cylindrical tube, the pressure can be obtained using a completely different approach, based on both  the asymptotic  interface profile shown in Fig. \ref{fig_asympt_profile} and on the static Laplace relation. This has been derived  by Bretherton  \cite{bretherton61} and in a more rigorous formalism by Park and Homsy \cite{park84}.

In the small capillary regime discussed in this paper, the bubble is almost at equilibrium far from the wall. Its shape is thus a spherical cap. As discussed in section \ref{parabole}, the extrapolation of the outside profile reaches its minimum at the position  $h_p= r_m (3 \beta Ca)^{2/3} H''_{f,out} H_p$ (positive or negative). The actual radius of curvature of the meniscus is thus slightly smaller than $r_t$ at the front and slightly larger at the rear side. At dominant order in $Ca$, the dynamical radius of curvature is thus  
\be
r_m^{f/r^*} = r_t \left(1+ (3 \beta Ca)^{2/3} H''_{f,out} H_p^{f/r^*} \right)
\ee

The Laplace pressure becomes:
\be
p^{f/r^*} = p^0 - \frac{2 \gamma(-\infty)}{r_m^{f/r^*} }  \sim  p^0 - \frac{2}{r_t}\left(\gamma(-\infty) + \gamma(3 \beta Ca)^{2/3} H''_{f,out} H_p^{f/r^*} \right) \; .
\label{pressure_bretherton}
\ee

This equation is the same as eq. \ref{p_drop} (with ${\cal P}/{\cal S}= 2/r_t$), but this equivalency requires some discussion.
In the stress free case, $\gamma$ is everywhere equals to $\gamma_{eq}$. We deduce from eq. \ref{f_v_gene} that  eq. \ref{p_drop} and \ref{pressure_bretherton} are identical only if 
\be
F^{f/r}= H''_{f,out}H_p^{f/r} \, .
\label{force_pression}
\ee
This relation can easily be proved by integration by parts of $F^{f/r}$, using the relation \ref{systeme}. 

%1.85=0.643*2.88   OK
%-0.53= 0.643*-0.82  OK

For incompressible interfaces, the situation is more subtle, as the correction in the surface tension is of the same order as the correction in the curvature. Thus both need to be taken into account in $\delta p$.  The precise value of $\gamma(-\infty)$ is given by $\sigma^{int}_{-\infty}$. Using, as previously, $\sigma^{int,f} (\ell) =\sigma^{int,r^*} (\ell)$,  $\delta p$ becomes
\be
\delta p=- \frac{2}{r_t} \left(\delta \sigma^{int,f} -\delta \sigma^{int,r^*} \right) - \frac{2 \gamma}{r_t}(3 \beta Ca)^{2/3} H''_{f,out} (H^{r^*}_p -H^f_p) \, . 
\label{dp_bretherton_inc}
\ee
Inserting eq. \ref{dgammma} and \ref{force_pression} into this latest expression leads to the same result as given by eq. \ref{dp_fv_sli} and \ref{dp_fv_rol}, thus insuring the consistency of both approaches.

The pressure correction induced by the surface tension variation is not considered in \cite{ratulowski90}, thus explaining the different pressure value proposed for the rolling case in Fig. 6 of \cite{ratulowski90}. 
It should also be noted that using a direct analogy of this method to obtain the pressure drop for a lamella, as made in  \cite{hirasaki85}, leads to a wrong relation between the friction force $f$ and the meniscus radius.

\section{Boundary conditions for foam flows}
\label{filmfoam}

\subsection{Film orientation}

\begin{figure}
\centerline{\includegraphics[width=6cm]{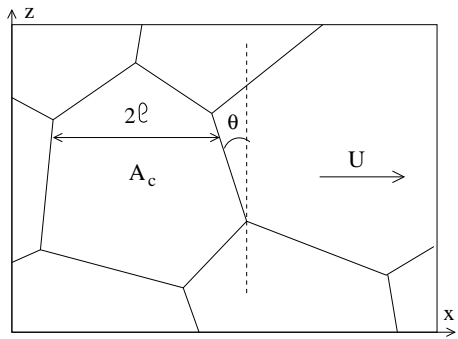} }
\caption{Network of menisci in contact with a wall, for a 2D or 3D dry foam. Each meniscus makes an angle $\theta$ with the direction $z$, normal to the foam velocity. The contact area between a bubble and the wall is $A_c$. }
\label{dry_foam}
\end{figure}

The friction between a  liquid foam and a smooth solid wall is governed by the force exerted on each meniscus in contact with the solid. The model previously developed only considers a meniscus moving in a direction perpendicular to its axis.  However, in a foam, the menisci in contact with the wall are oriented in all directions.
Comparison of the pressure drops associated with  lamellae of  different orientations and  moving in channel of rectangular section, suggests that only the projected length, in the direction perpendicular to the flow, contributes to the pressure drop  \cite{cantat04, raven06b,marmottant09}. However, this conjecture, that was natural for a foam having a motion of solid translation, is difficult to extrapolate in a general case. 
Indeed, during a complex deformation, only the normal velocity of the film is experimentally well defined. 
It is thus more consistent to assume, as in the viscous froth model \cite{kern04,grassia08}, that, locally, the force is oriented perpendicular to the meniscus and depends only on the normal velocity.
We consider a portion of meniscus of length $ds$, that makes an angle $\theta$ with the $x$ direction  and that moves with the apparent velocity  $U \vec{x}$ (see Fig. \ref{dry_foam}). 
The viscous forces $dF_x$ projected in the $x$ direction acting on this element are respectively, for the projected length model ($pl$) and for the non linear viscous force model ($vfm$),
\be
dF^{pl}_x =- f \cos \theta ds \, , 
\label{force_pl}
\ee
and
\be
dF^{vfm}_x =- \xi \left(\frac{\eta U}{\gamma} \right)^{2/3} (\cos \theta)^{5/3} ds -  \zeta \left(\frac{\eta U}{\gamma} \right)^{1/3} \frac{\ell}{r_m} (\cos \theta)^{4/3} ds \, , 
\label{force_vfm}
\ee
with $f$ the viscous force per unit length of meniscus as defined by eq.\ref{force_tot}.

No clear experimental evidence allows us to discriminate between these two models yet.   
For the sake of simplicity, we will use eq. \ref{force_pl} in the following section. 

\subsection{Influence of the liquid fraction}

The effect of the  liquid fraction is implicit in the sliding case, through the ratio $\ell/r_m$ (see eq. \ref{fsli}). 
Additionally,  the tangential stress between the foam and the wall depends on the total amount of
 contact lines between the meniscus and the wall, per unit surface, which strongly depends on the bubble volume  $V= 4 \pi r_b^3/3$ and on the liquid fraction  $\phi$ \cite{denkov05}.

\subsubsection{3D dry foams}

In the dry foam limit, the contact between a bubble and the wall is a polygon of area $A_c$ (see fig. \ref{dry_foam}). The statistical relation between the bubble radius $r_b$ and its contact area  has been investigated numerically in \cite{wang09}, and, for a monodisperse foam, the average contact area is found to be  $A_c=2.5 \, r_b^2$.  
Assuming a hexagonal shape of arbitrary orientation for the contact area, we get the averaged projected length of meniscus  per bubble, $d_{proj}$, as derived in \cite{raufaste09}:
\be
d_{proj}= 3 \langle |\cos \theta| \rangle \sqrt{\frac{2A_c}{3\sqrt{3}}} = 1.87 r_b
\label{proj}
\ee
%OK avec Raufaste 09 eq5 ( 1.87/2.5^0.5= 1.18)
Here we neglect the area occupied by the meniscus itself on the solid wall. For slightly larger liquid fractions a correction  has been proposed in \cite{raufaste09}.

If the interfaces are incompressible, the force $f$ depends on $r_m$ and $\langle \ell \rangle$, both of which must be determined as a function of $\phi$ and $r_b$.
In the limit of low liquid fraction, the radius of curvature of the meniscus is well approximated by 
$r_m=1.73 r_b \sqrt{\phi} $ \cite{livre_mousse}. Computing  the average distance between both sides of a regular hexagon (with 2 edges perpendicular to the velocity direction)  we obtain  $\langle \ell \rangle = 0.64 r_b$ and 
\be
\langle \ell \rangle/r_m = 0.37 \phi^{-1/2}\, .  
\label{lmoysec}
\ee
%0.64/1.73 = 0.37

Finally, for a monodisperse dry foam, the tangential stress exerted by the wall is 
\be
\sigma_{dry} = \frac{d_{proj}}{A_c} f(Ca,\langle \ell  \rangle /r_m) = \frac{0.74}{r_b} f(Ca, 0.37 \phi^{-1/2})   \, , 
\label{sigma_dry}
\ee
with the expression of the force per unit length of meniscus $f$ given by eq. \ref{force_tot}, for each kind of interface, and $r_b$ the bubble radius.

\subsubsection{Wet foams}

\begin{figure}
\centerline{\includegraphics[width=5cm]{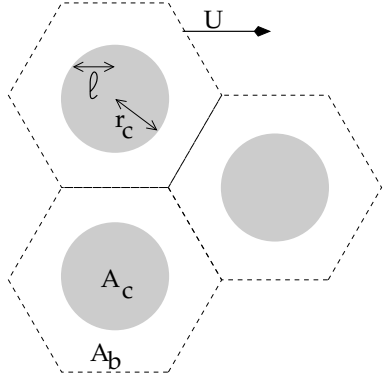} }
\caption{Wet foam in contact with a wall. Each bubble occupies a hexagonal domain of area $A_b=\pi r_b^2$. The contact area between the bubble and the wall is a disc of area $A_c= \pi r_c^2$ represented in grey. 
 }
\label{wet_foam}
\end{figure}

In a wet foam, the bubbles are much less deformed and their shape remains closer to a sphere. The area per bubble on the wall is thus close to $A_b= \pi r_b^2$. The contact area is a disc of area
$A_c= \pi r_c^2$, as depicted on Fig. \ref{wet_foam}.
The fraction of the wall $\alpha^2$  in contact with the bubbles has been measured by Princen \cite{princen85}:
\be
\alpha^2= \frac{r_c^2}{r_b^2}= 1-\frac{3.2}{\left(\frac{1-\phi}{\phi} +7.7\right)^{0.5}} \, .
\label{f_princen}
\ee
The total amount of  meniscus (with its rear and front part) per bubble, projected in the direction perpendicular to the motion, is  thus given  by 
$d_{proj}= 2 r_c = 2 r_b \alpha $ and the average wetting film length by
$ \langle \ell \rangle =  \frac{\pi}{4} \alpha r_b $.
Finally, following the approach proposed by Denkov {\it et al.} in \cite{denkov05}, the radius of the meniscus $r_m$ is determined  from its relation with the osmotic pressure of the foam $\Pi_{osm}$ \cite{princen86a}:
\be
\Pi_{osm} = \alpha^2 \gamma \, \kappa \, ,
\label{osm_pre}
\ee
with $\kappa$ the interface curvature, outside the contact area. In this wet regime the bubbles are close to spheres with flat patches, giving $\kappa \sim 2/r_m$. So we can express $r_m$ as $
r_m = 2 \gamma  \alpha^2/\Pi_{osm}$ and 
\
\be
\langle \ell \rangle/r_m = \frac{\pi r_b}{8 \alpha \gamma }  \Pi_{osm}\, .  
\label{lmoywet}
\ee

The osmotic pressures measured by Princen and Kiss \cite{princen87} are described by the phenomenological relation 
\be
\Pi_{osm,Pr}= \frac{\gamma}{r_b}\frac{0.00819 (1-\phi)^2}{(0.0361+0.9639 \phi)^2} \; .  
\label{piosm_princen} 
\ee 
H\"ohler {\it et al.} provided more recent numerical and experimental data for $\Pi_{osm}$ in \cite{hohler08}, approximated  by the slightly different relation
\be
\Pi_{osm, Ho}=7.3 \frac{\gamma}{r_b} \frac{(0.26-\phi)^2}{\sqrt{\phi}}\; .  
\label{piosm_hohler} 
\ee

These relations enable us to predict the tangential stress at the wall as a function of the bubble radius $r_b$ and the liquid fraction $\phi$ in a wet foam:

\be
\sigma_{wet} = \frac{d_{proj}}{A_c} f(Ca,\langle \ell \rangle /r_m) =\frac {2  \alpha}{\pi r_b}  f(Ca,\langle \ell \rangle /r_m)\, , 
\label{sigma_wet}
\ee
with $\alpha$ and $\langle \ell \rangle/r_m$ expressed as function of $\phi$ and $r_b$ in eq. \ref{f_princen}   and \ref{lmoywet}, and $f$ given by eq. \ref{force_tot}.

\subsection{Comparison with some experimental results}

\subsubsection{3D Foams}

Foam rheological properties are usually investigated with a rheometer in cone/plane geometry. 
Using smooth plates, there is a velocity range in which the foam is not sheared in bulk and only slips on the bottom plate. The relation between the slip velocity and the 
stress can thus be deduced, as a function of the bubble size and liquid fraction.

For foaming solutions leading to stress free interfaces, the scaling $\sigma \sim Ca^{2/3}$ is well verified \cite{denkov05}. However, the theoretical law obtained from eq. \ref{sigma_wet} and eq. \ref{force_tot} must be corrected by a prefactor close to 4. Similarly high prefactors  have also been observed for 2D foam \cite{raufaste09} and lamellae \cite{cantat04,dollet10}. \\

\begin{figure}[h]
\centerline{\includegraphics[width=8cm]{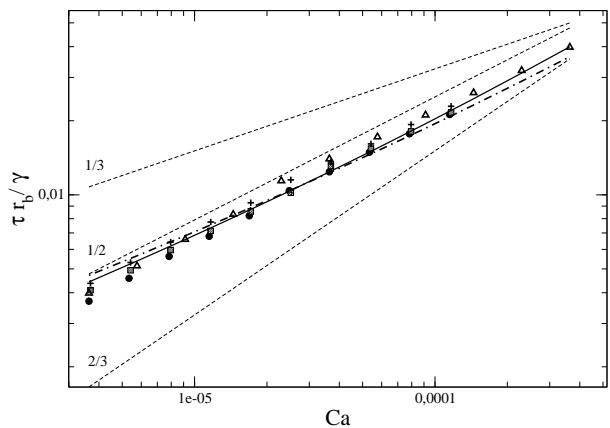} }
\caption{Stress at the wall rescaled by $\gamma/r_b$ as a function of the capillary number $Ca$ for a 3D foam. Symbols: Experimental data obtained by Denkov {\it et al.} in \cite{denkov05}, Fig. 8c. The solid line is obtained from eq. \ref{f_mousse} with $Z= 0.7$ and $\ell/r_m=0.318$, and the dot-dashed line with  $Z= 0.5$ and $\ell/r_m=0.5$. The dashed lines are respectively power laws of exponent 1/3, 1/2 and 2/3.}
\label{fig_friction_denkov}
\end{figure}

 Experimental data have been obtained by Denkov {\it et al.} for several foaming solutions of  very high interfacial dilatational  modulus \cite{denkov05}. The results of Section \ref{section_profile_denkov} show that the film profiles obtained with this kind of foaming solution are well fitted by the sliding model of interface. We thus compare the experimental  stress  with the prediction from eq. \ref{sigma_wet}, using the expression of $f$ corresponding to the sliding interface (eq. \ref{force_tot}):
  \be
\sigma_{wet,sli}= \frac{\gamma}{r_b} Z \left[6.82\, Ca^{2/3}+0.94 \frac{\langle \ell \rangle}{r_m}Ca^{1/3}\right ]
\label{f_mousse}
\ee
with $Z= 2  \alpha/\pi$.  
The experimental liquid fraction is $\phi= 0.1$. The two models of osmotic pressure eq. \ref{piosm_princen} and eq.  \ref{piosm_hohler}  lead respectively to $\left.\langle \ell \rangle /r_m \right)^{Pr}= 0.318$  and  $\left.\langle \ell \rangle/r_m \right)^{Ho}= 0.5$. 
Both predictions have been plotted on Fig. \ref{fig_friction_denkov},  keeping  $Z$ as an adjustable parameter. A very good agreement 
is obtained with  $Z^{fit,Pr}=0.7$ and $Z^{fit,Ho}=0.5$, whereas the theoretical prediction for the prefactor  is  $Z^{theo}=0.3$. 
Consistently with the observations at the single meniscus scale, the model slightly under-predicts the force. 
However, the variation with the capillary number is quantitatively reproduced by our model, without adjustable parameter on $\langle \ell \rangle /r_m$.

\vspace*{1cm}

\begin{figure}[h]
\centerline{\includegraphics[width=8cm]{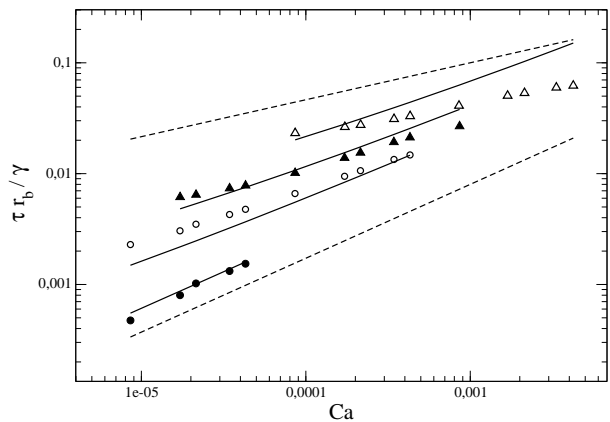} }
\caption{Stress at the wall, rescaled by $\gamma/r_b$, plotted  as a function of the capillary number for a  3D foam of Amilite GCK-12. $\vartriangle,\blacktriangle, \circ, \bullet$: Experimental data from Fig.8 in \cite{marze08}, with liquid fractions $\phi$=0.10, 0.15, 0.20 and 0.25,  respectively.  The solid lines are predictions of eq. \ref{f_mousse} using the theoretical values of  $\langle \ell \rangle/r_m =0.5; 0.25; 0.09$ and 0.004, as obtained from eq. \ref{lmoywet} and  eq. \ref{piosm_hohler} and $Z^{fit}= 2\, Z^{theo}=0.6; 0.44; 0.32$ and $0.18$. The dashed lines are power laws of exponent 1/3 and 2/3.}
\label{fig_friction_arnaud}
\end{figure}

Another set of experimental data obtained  by Marze {\it et al.}, with a solution of Amilite GCK-12,  tests the variation of the stress with the liquid fraction in the range $\phi=0.1 - 0.25$ \cite{marze08}. The data, plotted as a function of the capillary number,  are well fitted by power laws, with an exponent $n$ that depends on the liquid fraction: $n=(0.28; 0.44; 0.48; 0.66)$ for $\phi= (0.1; 0.15; 0.20; 0.25)$. Qualitatively, the exponent varies continuously  from $2/3$ for a dynamic  dominated by the meniscus (high liquid fraction) to $1/3$ for a dynamic  dominated by the wetting film (moderate liquid fraction), as expected for incompressible interfaces in the sliding regime. 
A quantitative comparison of these data with the prediction of eq. \ref{f_mousse}  is shown in Fig. \ref{fig_friction_arnaud} using the expression  eq. \ref{piosm_hohler} for  the osmotic pressure, as it gives a much better prediction for high liquid fractions than eq. \ref{piosm_princen}.
The agreement is very good for the  highest liquid fractions but the $1/3$ power law obtained for the smallest value of $\phi$ is not reproduced. 
In that case, the best agreement is obtained for  $Z^{fit} = 2.0 Z^{theo}$  (using a single adjustable parameter for the four data series).  

\vspace*{1cm}

\subsubsection{2D Foams}

The rheological properties of a foam strongly depends on its organization at the bubble scale and the need to observe this local structure is at the origin of the numerous experiments on 2D foams. The bubbles are organized in a single monolayer that is confined either between two glass plates, between the solution and a glass plate or between the solution and the atmosphere. In the first two cases, the viscous friction at the plate plays an important role in the dynamics and is often several order of magnitude higher than the internal viscous force. 
The stress at the plates has been measured by Raufaste {\it et al.} as a function of the capillary number, the meniscus size and the bubble area \cite{raufaste09}. 
All their experimental data are well described by the phenomenological law
\be
 f=5.13 \, \gamma \, Ca^{2/3} \left(\frac{\sqrt{A_c}}{r_m} \right)^{0.48}
\ee
%38\times 3.72
with $A_c$ the area per bubble.
The velocity power law in 2/3 for the force per unit length of meniscus is not consistent with a dependency on the meniscus radius. 
However, fitting only the velocity dependency, they found $f\sim Ca^{0.58}$ which is between the $1/3$ and $2/3$ exponents. 
A finite interfacial  compressibility may  thus be at the origin of this behavior.

\section{Conclusion}
\label{conclusion}
In this paper, we show that a part of the available experimental results on meniscus motion can be explained using very simple assumptions for the interfacial properties. 
We focus on both the viscous force acting on the moving meniscus  and on the induced  pressure drop. These  are quantities of high practical interest for controlling or predicting bubbles or foam motion close to solid walls. In particular, we establish new predictions for these  quantities in the limit of incompressible interfaces.
However, wetting film thickness measurements clearly evidence intermediate regimes, between the stress-free interface and the incompressible one \cite{stebe91,shen02,delacotte12,cantat12}, and the influence of these  on the forces are still unclear. The surfactant concentration at the interface is governed by convective and diffusive processes, as well as adsorption and desorption rates. The interfacial stress is modified by the surfactant's intrinsic shear and dilatational viscosities and by the variation of the surface tension. 
More realistic models, taking into account some or all of these interfacial properties, are developed numerically  \cite{hirasaki85,ratulowski90,stebe91,park92,stebe95,ramdane97,breward02,shen02,ghadiali03,scheid10}. Nevertheless, the rheological properties of the interfaces are not very well characterized at the high extensional rates generated by the meniscus motion and it is often difficult to identify the dominating processes. 
Moreover, for incompressible or almost incompressible interfaces, the outer flow, far from the dynamical meniscus, may be strongly modified by the moving interface \cite{mayer12}. 
The relevant boundary conditions required to model the interface motion are thus difficult to identify. Different assumptions for the outer flow may lead, in the limit of high surface modulus, to either the rolling  or  the sliding limiting cases. Consequently,  solving the hydrodynamical problem at a larger scale, and not only in the dynamical meniscus domain, is probably unavoidable.

\begin{acknowledgments}
We thank  N. Denkov and S. Tcholakova for their experimental data and  A. Saint-jalmes, B. Dollet, J. Seiwert,  S. Jones and H. A. Stone for their constructive remarks. 
\end{acknowledgments}

%\bibliographystyle{/home/cantat/isabelle/bib/report}
%\bibliography{/home/cantat/isabelle/bib/bib}

\end{document}